\documentclass[prd,a4paper,twocolumn,preprintnumbers,floatfix]{revtex4}
\usepackage{graphicx}

\newcommand{\nc}{\newcommand}

\nc{\be}[1]{\begin{equation}\mbox{$\label{#1}$}}
\nc{\bea}[1]{\begin{eqnarray} \mbox{$\label{#1}$}}
\nc{\Section}[2]{\section{#2}\label{#1}}
\nc{\Bibitem}[1]{\bibitem{#1}}
\nc{\Label}[1]{\label{#1}}

\nc{\eea}{\end{eqnarray}}
\nc{\ee}{\end{equation}}

\nc{\bdm}{\begin{displaymath}}
\nc{\edm}{\end{displaymath}}
\nc{\dpsty}{\displaystyle}
\nc{\bc}{\begin{center}}
\nc{\ec}{\end{center}}
\nc{\ba}{\begin{array}}
\nc{\ea}{\end{array}}
\nc{\bab}{\begin{abstract}}
\nc{\eab}{\end{abstract}}
\nc{\btab}{\begin{tabular}}
\nc{\etab}{\end{tabular}}
\nc{\bit}{\begin{itemize}}
\nc{\eit}{\end{itemize}}
\nc{\ben}{\begin{enumerate}}
\nc{\een}{\end{enumerate}}
\nc{\bfig}{\begin{figure}}
\nc{\efig}{\end{figure}}

\nc{\arreq}{&\!=\!&}
\nc{\arrmi}{&\!-\!&}
\nc{\arrpl}{&\!+\!&}
\nc{\arrap}{&\!\!\!\approx\!\!\!&}
\nc{\non}{\nonumber}
\nc{\align}{\!\!\!\!\!\!\!\!&&}

\def\lsim{\; \raise0.3ex\hbox{$<$\kern-0.75em
      \raise-1.1ex\hbox{$\sim$}}\; }
\def\gsim{\; \raise0.3ex\hbox{$>$\kern-0.75em
      \raise-1.1ex\hbox{$\sim$}}\; }

\nc{\DOT}{\hspace{-0.08in}{\bf .}\hspace{0.1in}}
\nc{\Laada}{\hbox {$\sqcap$ \kern -1em $\sqcup$}}
\nc\loota{{\scriptstyle\sqcap\kern-0.55em\hbox{$\scriptstyle\sqcup$}}}
\nc\Loota{{\sqcap\kern-0.65em\hbox{$\sqcup$}}}
\nc\laada{\Loota}
\nc{\qed}{\hskip 3em \hbox{\BOX} \vskip 2ex}

\nc{\real}{{\rm I \! R}}
\nc{\Z}{{\sf Z \!\!\! Z}}
\nc{\complex}{{\rm C\!\!\! {\sf I}\,\,}}
\def\bigid{\leavevmode\hbox{\small1\kern-3.8pt\normalsize1}}
\def\id{\leavevmode\hbox{\small1\kern-3.3pt\normalsize1}}
\nc{\slask}{\!\!\!/}
\nc{\bis}{{\prime\prime}}
\nc{\pa}{\partial}
\nc{\na}{\nabla}
\nc{\ra}{\rangle}
\nc{\la}{\langle}
\nc{\goto}{\rightarrow}
\nc{\swap}{\leftrightarrow}

\nc{\EE}[1]{ \mbox{$\cdot10^{#1}$} }
\nc{\abs}[1]{\left|#1\right|}
\nc{\at}[2]{\left.#1\right|_{#2}}
\nc{\norm}[1]{\|#1\|}
\nc{\abscut}[2]{\Abs{#1}_{\scriptscriptstyle#2}}
\nc{\vek}[1]{{\rm\bf #1}}
\nc{\integral}[2]{\int\limits_{#1}^{#2}}
\nc{\inv}[1]{\frac{1}{#1}}
\nc{\dd}[2]{{{\partial #1}\over{\partial #2}}}
\nc{\ddd}[2]{{{{\partial}^2 #1}\over{\partial {#2}^2}}}
\nc{\dddd}[3]{{{{\partial}^2 #1}\over
    {\partial #2 \partial #3}}}
\nc{\dder}[2]{{{d #1}\over{d #2}}}
\nc{\ddder}[2]{{{d^2 #1}\over{d {#2}^2}}}
\nc{\dddder}[3]{{d^2 #1}\over
    {d #2 d #3}}
\nc{\dx}[1]{d\,^{#1}x}
\nc{\dy}[1]{d\,^{#1}y}
\nc{\dz}[1]{d\,^{#1}z}
\nc{\dl}[1]{\frac{d\,^{#1}l}{(2\pi)^{#1}}}
\nc{\dk}[1]{\frac{d\,^{#1}k}{(2\pi)^{#1}}}
\nc{\dq}[1]{\frac{d\,^{#1}q}{(2\pi)^{#1}}}

\nc{\bfT}{{\bf T }}

\nc{\cA}{{\cal A}}
\nc{\cB}{{\cal B}}
\nc{\cD}{{\cal D}}
\nc{\cE}{{\cal E}}
\nc{\cG}{{\cal G}}
\nc{\cH}{{\cal H}}
\nc{\cL}{{\cal L}}
\nc{\cO}{{\cal O}}
\nc{\cT}{{\cal T}}
\nc{\cN}{{\cal N}}
\nc{\cR}{{\cal R}}
\nc{\rvac}[1]{|{\cal O}#1\rangle}
\nc{\lvac}[1]{\langle{\cal O}#1|}
\nc{\rvacb}[1]{|{\cal O}_\beta #1\rangle}
\nc{\lvacb}[1]{\langle{\cal O}_\beta #1 |}
\nc{\bb}{\bar{\beta}}
\nc{\bt}{\tilde{\beta}}
\nc{\ctH}{\tilde{\cal H}}
\nc{\chH}{\hat{\cal H}}
\nc{\1}{\aa}
\nc{\2}{\"{a}}
\nc{\3}{\"{o}}
\nc{\4}{\AA}
\nc{\5}{\"{A}}
\nc{\6}{\"{O}}
\nc{\al}{\alpha}
\nc{\g}{\gamma}
\nc{\Del}{\Delta}
\nc{\e}{\textrm{e}}
\nc{\eps}{\epsilon}
\nc{\lam}{\lambda}
\nc{\Om}{\Omega}
\nc{\ve}{\varepsilon}
\nc{\mn}{{\mu\nu}}
\nc{\vp}{\varphi}

\nc{\rf}[1]{(\ref{#1})}
\nc{\nn}{\nonumber \\*}
\nc{\bfB}{\bf{B}}
\nc{\bfv}{\bf{v}}
\nc{\bfx}{\bf{x}}
\nc{\bfy}{\bf{y}}
\nc{\vx}{\vec{x}}
\nc{\vy}{\vec{y}}
\nc{\oB}{\overline{B}}
\nc{\oI}{\overline{I}}
\nc{\oR}{\overline{R}}
\nc{\rar}{\rightarrow}
\nc{\ti}{\times}
\nc{\slsh}{\hskip-5pt/}
\nc{\sm}{Standard~Model~}
\nc{\MP}{M_{\rm Pl}}
\nc{\mpl}{M_{\rm Pl}}
\nc{\tp}{t_{\rm Pl}}

\nc{\pmin}{p_{\rm min}}
\nc{\pmax}{p_{\rm max}}
\nc{\fo}{f_0}
\nc{\foi}{f_{0,i}\,}
\nc{\fop}{f_0^P}
\nc{\fou}{f_0^U}

\nc{\eff}{{\rm eff}}
\nc{\MT}{M_{\rm T}}
\nc{\ML}{M_{\rm L}}
\nc{\kk}{\vek{k}}
\nc{\pp}{{\rm p}}
\nc{\pt}{\partial_t}
\nc{\half}{{1\over 2}}
\nc{\w}{\omega}
\nc{\uhat}{\hat{U}_\w}

\nc{\etal}{\mbox{\it et al.}}
\nc{\ie}{{\it i.e. }}
\nc{\eg}{{\it e.g. }}
\nc{\trh}{T_{\rm RH}}
\nc{\ad}{{a'\over a}}
\nc{\bd}{{b'\over b}}
\nc{\Rd}{{R'\over R}}
\nc{\diag}{{\textrm{diag}}}
\nc{\mato}[1]{\tilde{#1}}
\nc{\sech}{\textrm{sech}}
\nc{\I}{\textrm{I}}
\nc{\II}{\textrm{II}}
\nc{\III}{\textrm{III}}
\nc{\vev}[1]{\langle #1 \rangle}
\nc{\hyp}{\,\; F_{1{\hskip -16pt}2}{\hskip 11pt}}
\nc{\brhom}{\overline{\rho}_M}
\nc{\brho}{\overline{\rho}}
\nc{\rhob}{\overline{\rho}}
\nc{\Pb}{\overline{P}}
\nc{\bH}{\overline{H}}
\nc{\ep}{{1+4\eps}}

\def\smiley{\hbox{\large$\bigcirc$\hspace{-.80em}%
\raise.2ex\hbox{$\cdot\cdot$}\kern-.61em    
\lower.2ex\hbox{\scriptsize$\smile$}}\ }

\def\frowney{\hbox{\large$\bigcirc$\hspace{-.80em}%
\raise.2ex\hbox{$\cdot\cdot$}\kern-.635em
\lower.2ex\hbox{\scriptsize$\frown$}}\ }

\nc{\om}{\Omega_{\textrm{\footnotesize m}}}
\nc{\oce}{\Omega_{\textrm{\footnotesize C}}}
\nc{\olam}{\Omega_{\footnotesize\Lambda_{\textrm{b}}}}
\nc{\oll}{\Omega_{\footnotesize \ell}}
\nc{\osig}{\Omega_{\footnotesize \sigma}}

\nc{\lcdm}{$\Lambda$CDM }

\begin{document}
\title{Loitering universe models in light of the CMB}
\author{\O. Elgar\o y}
\affiliation{Institute of theoretical astrophysics, University of Oslo, Box 1029, 0315 Oslo, Norway}
\affiliation{NORDITA, Blegdamsvej 17, DK-2100, Copenhagen, Denmark}
\author{D. F. Mota}
\affiliation{Institute of theoretical astrophysics, University of Oslo, Box 1029, 0315 Oslo, Norway}
\affiliation{ Department of Physics, University of Oxford, Keble Road, 
Oxford OX1 3RH, UK}
\author{T. Multam\"aki}
\affiliation{NORDITA, Blegdamsvej 17, DK-2100, Copenhagen, Denmark}
\date{\today}

\preprint{NORDITA-2005-5}

\begin{abstract}
Spatially flat loitering universe models have recently been shown to arise in the context of 
brane world scenarios. Such models allow more time for structure formation 
to take place at high redshifts, easing, e.g., the tension between the observed and predicted evolution of the quasar 
population with redshift.  
While having the desirable effect of boosting the growth of
structures, we show that in such models the position of the 
first peak in the power spectrum of the cosmic microwave background 
anisotropies severely constrain the amount of loitering at high redshifts.

\end{abstract}
\maketitle

\section{Introduction}

The last few years of intense observational developments have provided 
cosmology with a standard model: a universe with geometry described by the 
flat Robertson-Walker metric, with cold pressureless matter contributing 
roughly 1/3 and some form of negative pressure `dark energy' contributing 
the remaining 2/3 of the critical energy density.  The existence of the 
last component is motivated partly by the Hubble diagram from supernovae 
of type Ia (SNIa) \cite{riess,perlmutter}, 
partly from joint analysis of the cosmic microwave background (CMB) 
anisotropies and the large-scale distribution of matter, e.g. from 
the power spectrum of the galaxy distribution \cite{efstathiou,tegmark}.  
However, there are 
ways of describing the data which do not invoke a negative-pressure 
fluid, namely to postulate some modification of standard Einstein 
gravity on large scales.  One way of realizing this is in the 
braneworld scenario, where our universe is taken to be a (3+1)-dimensional 
membrane (the brane) residing in a higher-dimensional space (the bulk), 
see \cite{shtanov} for an overview.    
The standard model fields are confined to the brane, whereas gravity can 
propagate in the full space.  The extra dimensions need not be small, and 
hence gravity can be modified on scales as large as the size of the 
present horizon $\sim c/H_0\sim 3000\;h^{-1}\,{\rm Mpc}$ (where 
$c$ is the speed of light, set equal to 1 in the remainder of this paper, 
and $H_0=100h\;{\rm km},\,{\rm s}^{-1}\,{\rm Mpc}^{-1}$ is the present 
value of the Hubble parameter.)  

Loitering, i.e., a universe which undergoes a phase of very slow 
growth, can arise in closed universe models with a cosmological 
constant \cite{sahni2}.  
In the context of extra-dimensional models  such phenomena 
can occur naturally in spatially flat geometries and may provide a solution to
 several problems. 
For instance, in brane gas cosmology(BGC) \cite{bgc}, 
loitering may help to solve the 
horizon problem and the brane problem of BGC \cite{brand}.
Recently, the possibility of a loitering 
phase has been pointed out by Sahni and Shtanov \cite{sahni} in the 
context of other braneworld models, where the geometry on our brane is 
spatially flat.  Such a phase has desirable consequences, since it 
allows more time for structure formation \cite{feldman} and 
astrophysical processes, alleviating some of the tension 
between the concordance $\Lambda{\rm CDM}$ model and the observations of 
quasars with redshifts $z\sim 6$ \cite{richards,haiman} and the 
indications from WMAP of early reionization \cite{wmap}.  
The loitering phase is proposed to occur at high redshifts $z\sim 20$, and   
the behavior of these loitering models at moderate redshifts is 
similar to the $\Lambda{\rm CDM}$ model, thus making it possible to satisfy 
constraints from e.g. SNIa.  However, as we will show in this paper, 
the position of the first peak of the power spectrum of the CMB 
anisotropies places a very robust constraint on the high-redshift behavior 
of all models.  

The structure of this Brief Report is as follows.  We start out by 
introducing the loitering braneworld models in section II, and 
confront them with data in section III.  In section IV we look at 
constraints on loitering in general from the CMB peak positions. 
We summarize and conclude in section V.  

\section{The loitering brane world model}

The braneworld models considered in \cite{sahni} are defined by the action 
\bea{action}
S & = & M^3\Big[\int_{\rm bulk} d^5x\sqrt{-g}\Big(\mathcal{R}-2\Lambda_b\Big)
-2\int_{\rm brane}^{}d^4x\sqrt{-g^{(4)}} K\Big]\nonumber\\
& & +m^2\int_{\rm brane}^{}d^4x\sqrt{-g^{(4)}}\Big(\mathcal{R}^{(4)}
-2{\sigma\over m^2}\Big)\nonumber\\
& & +\int_{\rm brane}^{}d^4x\sqrt{-g^{(4)}}\mathcal{L},
\eea
where $M$ and $m$ are the five and four dimensional Planck masses,
$\Lambda_b$ the bulk cosmological constant and $\sigma$ the 
brane tension.
The Friedmann equation of the loitering brane world model is \cite{sahni}:
\bea{bwfried}
\Big({H(z)\over H_0}\Big)^2 & = & \om (1+z)^{3}+\osig+2 \oll
-2\sqrt{\oll}\Big[\osig+\oll\nonumber\\
& & +\om (1+z)^3+\olam+\oce (1+z)^4\Big]^{1/2}
\eea
where
\bea{defs}
 & & \om={\rho_{\rm m}^0\over 3m^2H_0^2},\ \osig={\sigma\over 3m^2H_02^2},\ 
\oll={1\over l^2H_0^2} \nonumber\\
& & \olam=-{\Lambda_b\over 6 H_0^2},\ \oce=-{C\over a_0^4H_0^2}.
\eea
The $\oce$ term is the dark radiation energy density arising
from the brane-bulk interaction.
The condition for a spatially flat universe is 
\be{sig}
\osig=1-\om+2\sqrt{\oll}\sqrt{1+\olam+\oce}.
\ee

Late time acceleration occurs at 
the critical length scale corresponding to the present horizon scale,
i.e., $l=2m^2/M^3\sim H_0^{-1}$, which sets $\oll\sim \cO(1)$.
Since $\olam$ corresponds to the five-dimensional bulk cosmological constant,
it can naturally have a much larger value than the other parameters 
that correspond
to quantities on the brane, $\olam\gg \om,\ \oce,\ \oll$.
Hence, at high redshifts, $z\gsim 10$,  at which loitering occurs,
we can well approximate Eq. (\ref{bwfried}) by:
\bea{apfried}
\Big({H(z)\over H_0}\Big)^2 & \approx & \om (1+z)^{3}+2\sqrt{\oll\olam}\nonumber\\
& & -2\sqrt{\oll}\Big(\olam+\oce (1+z)^4\Big)^{1/2}.
\eea
Note that this is somewhat different what was considered
in \cite{sahni} where they also drop the $\olam$ term inside
the last square root. However, using the example parameter values 
given in \cite{sahni}, one sees that this is not well justified.

As discussed in \cite{sahni}, the Friedmann equation can exhibit different
behavior depending on the values of the parameters. Here we are interested in 
loitering behavior with respect to the \lcdm model and hence instead 
of $H(z)$, we consider $X(z)\equiv H/H_{\Lambda{\rm CDM}}$. This function 
has a well defined minimum for parameter values for which the 
interpretation of $H(z)$ is not as straightforward  
(see Fig. 1 in \cite{sahni}). 

We can now unambiguously define the loitering redshift as $X'(z_{\rm loit})=0$.
In the high redshift approximation, one finds that 
\be{zloit}
1+z_{\rm loit}=\Big(3{\olam\over\oce}\Big)^{1/4}.
\ee
Positivity at this minimum point, $H^2(z_{\rm loit})>0$, gives us a condition 
on $\olam$:
\be{lamcond}
\olam> {2^4\over 3^3} {\oll^2\oce^3\over\om^4}.
\ee
Within the high loitering redshift approximation, it is then clear that for a given
loitering redshift, $z_{\rm loit}$, $\oce$ is constrained by
\be{occonst}
3{\olam\over (1+z_{\rm loit})^4}<\oce<({3^3\om^4\over 2^4\oll^2}\olam)^{1/3},
\ee
indicating that both $\oce$ and $\olam$ have a maximum value for a given
$z_{\rm loit}$.

\section{Constraints on braneworld loitering}  

The CMB shift parameter determines the shift of the peaks in the 
CMB power spectrum when cosmological parameters are varied  
\cite{bond,melchiorri,odman}.  It is given by 
\begin{equation}
{{\cal R}} = \sqrt{\Omega_{\rm m}}H_0r(z_{\rm dec}),
\label{eq:shiftparam}
\end{equation}
where $r(z)=\int_0^z dz/H(z)$ is the comoving distance in a flat 
universe, and $z_{\rm dec}$ is the redshift at decoupling.  
This expression is derived in the $\Lambda{\rm CDM}$ model, and  
depends on the ratio between the sound horizon at decoupling 
and the angular diameter distance to $z_{\rm dec}$.  We can 
apply it straightforwardly in our case also, since the only way 
equation (\ref{eq:shiftparam}) could be radically changed is if the 
sound horizon were to change significantly in the brane world models 
conidered here, and we have checked that this is not the case. 
Observationally, from WMAP we have $z_{\rm dec}=1088^{+1}_{-2}$, 
and ${{\cal R}}_{\rm obs} =1.716\pm 0.062$ \cite{wmap}.  

\begin{figure}[ht]
\includegraphics[width=8cm]{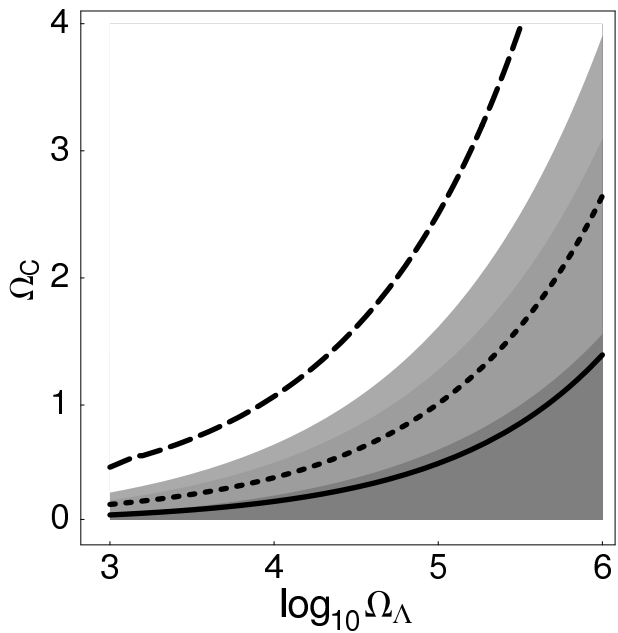}
\caption{Confidence contours (68(darkest), 95 and 99(lightest) \%) imposed by the 
CMB shift parameter on the parameters $\Omega_{\Lambda_{\rm b}}$ and $\Omega_{\rm C}$.
Also shown are lines of constant increase in age at $z=6$ compared with the 
$\Lambda{\rm CDM}$ model: $\Delta t=0.05$ (solid), $0.1$ (dotted), $0.3$ (dashed) Gyr.}
\label{fig:contour1}
\end{figure}

The WMAP constraint on ${{\cal R}}$ places severe constraints on 
loitering models.  
We have run a grid of models, fixing $\Omega_{\rm m}=0.3$, 
$\Omega_{\rm l}=3.0$, and allowing $\Omega_{\Lambda_{\rm b}}$ and 
$\Omega_{\rm C}$ to vary.   We compute ${{\cal R}}$ from Eq. 
(\ref{eq:shiftparam}), and compute $\chi^2 = ({{\cal R}}-{{\cal R}}_{\rm obs})
^2/\sigma_{{\cal R}}^2$, where $\sigma_{{\cal R}}=0.062$.  
The resulting contours are shown in Fig. \ref{fig:contour1}.  They 
effectively rule out any significant amount of loitering.  
For example, picking a model along the 68 \% contour, 
one finds that deviations 
from the standard $\Lambda{\rm CDM}$ behavior are tiny.  
In figure \ref{fig:contour1} we have also plotted contours of constant 
increase in age at $z=6$ for the loitering model compared with the 
$\Lambda{\rm CDM}$ model, in units of Gyr.  The age of the $\Lambda{\rm CDM}$ 
model at $z=6$ is $0.92\;{\rm Gyr}$ for $\Omega_{\rm m}=0.3$, $h=0.7$.  
Only a modest increase in age is allowed, a change of $0.3\;{\rm Gyr}$ being 
ruled out at more than 99\% confidence. 

\begin{figure}[ht]
\includegraphics[width=8cm]{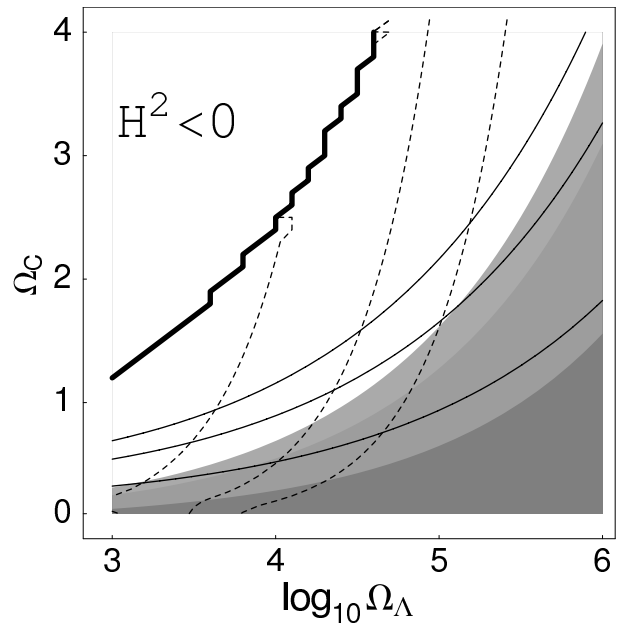}
\caption{The loitering redshift $z_{\rm loit}=10,\, 15,\, 20$ (dashed lines, left to right),
relative linear growth factor at $z=6$, $D/D_{\Lambda}=1.5,\, 2.0,\, 2.5$ 
(solid lines, bottom to top) and the shift parameter confidence regions (in gray)
for different values of $(\Omega_{\Lambda_{\rm b}},\Omega_{\rm C})$. Also shown is 
the excluded region where $H(z_{\rm loit})^2<0$.
}
\label{fig:contour2}
\end{figure}

One can also quantify the amount of loitering allowed by considering
how structures grow in the braneworld model considered here, compared to the \lcdm model.
The linear growth factor, $D$, is given by
\be{lingrow}
\ddot{D}+2H\dot{D}-\frac 32 {\om\over a^3} D=0
\ee
and it is easy to see that in the Einstein de Sitter-case, 
a simple growing solution exists, $D\sim a$.
We have calculated how the linear growth factor evolves in the braneworld model for
different values of the parameters and compared it to the value in the \lcdm model.
The results are shown in Fig. {\ref{fig:contour2}} for $z=6$. As a reference, for 
$\oce=8.0$, $\olam=4.5\times 10^{5}$ considered in \cite{sahni} (and clearly ruled out
by the shift parameter) for which 
$H(z)$ has a clear loitering phase at $z\sim 20$, one finds that
$D/D_{\Lambda CDM}(z=6)\approx 42$.
From the figure one sees that within the $99\%$ region, with $z_{\rm loit}=20$, 
the linear growth factor can only be about twice the value of that in the \lcdm model.

\section{Loitering in general}

The constraints found in the previous section are easily understood to 
arise from the fact that if the Hubble parameter is decreased compared 
to $\Lambda{\rm CDM}$ at high redshifts, the 
comoving distance to the last scattering surface is increased.  
Thus, the conclusion that loitering is effectively constrained by the 
CMB shift parameter is not specific to the braneworld model, 
and can be generalized as shown in the following.  

\begin{figure}[ht]
\includegraphics[width=6cm]{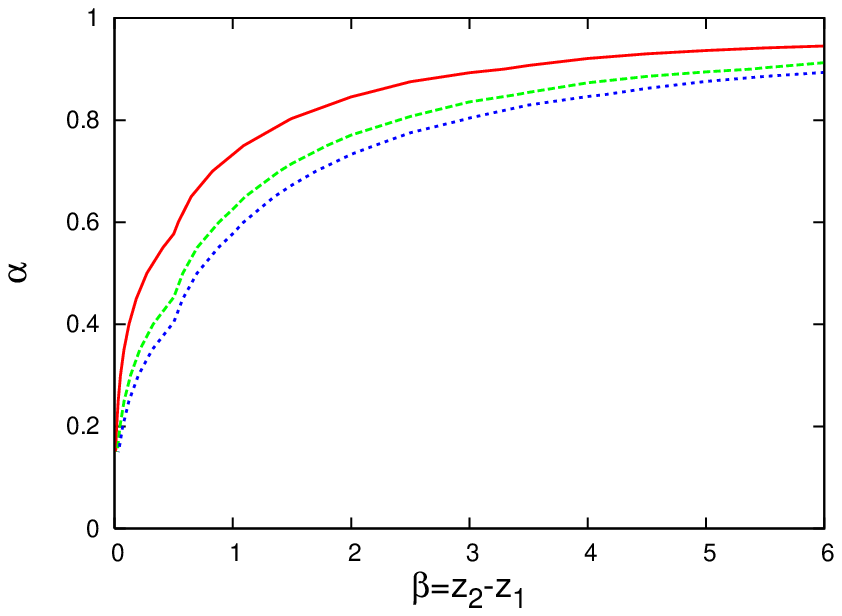}
\includegraphics[width=6cm]{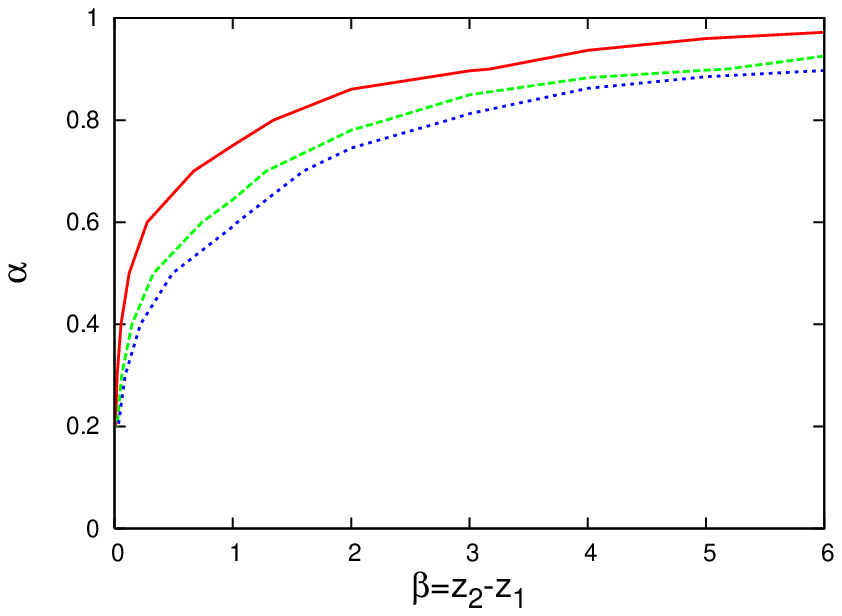}
  \caption{(Colour online only) 68, 95 and 99 percent confidence contours imposed by the 
CMB shift parameter on the parameters $\alpha$ and $\beta$ for $z_1=10$ 
(upper panel) and $z_1=500$ (lower panel). The $\Lambda$CDM model corresponds to the lines 
$\alpha=1$ or $\beta=0$ }
\label{fig:genloit}
\end{figure}

What makes a loitering phase attractive, is the fact that $H(z)$ is 
less than the Hubble parameter in $\Lambda{\rm CDM}$ during loitering, 
so that the universe can spend more time at high redshifts.  
In order to quantify this effect, we model the loitering phase by 
having a Hubble parameter that reduced by a factor $0<\alpha<1$ 
compared with its $\Lambda{\rm CDM}$ 
value in an interval $z_1 < z < z_2$, where $z_2 < z_{\rm dec}$.
Then the time the universe spends between $z_1$ and $z_2$ is increased 
by a factor $1/\alpha$, since $t(z_1)-t(z_2)=\int_{z_1}^{z_2} 
dz/((1+z)H(z))$.  
The corresponding change in the comoving distance to the last scattering 
surface compared with the $\Lambda{\rm CDM}$ model is 
$(1/\alpha -1)\int_{z_1}^{z_2}dz/H_{\Lambda{\rm CDM}}(z)$, and so 
the change in the shift parameter is 
\begin{equation}
\Delta{{\cal R}}=\sqrt{\Omega_{\rm m}}H_0\left(\frac{1}{\alpha}-1\right)
\int_{z_1}^{z_2}\frac{dz}{\sqrt{\Omega_{\rm m}(1+z)^3+(1-\Omega_{\rm m})}}.
\label{eq:est1}
\end{equation}
In Fig.  \ref{fig:genloit}, where we show the result of fitting the parameters 
$\alpha$ and $\beta\equiv z_2-z_1$ to the CMB shift parameter.  The upper panel shows 
the likelihood contours for a dip in the Hubble parameter which starts at 
a redshift of 10, and in the lower panel the dip starts at a redshift of 500. 
As can be seen from the figure, one can have either a substantial dip ($\alpha\rightarrow 0$) over 
a very small redshift interval ($\beta\approx 0$), or a small dip ($\alpha\approx 1$) 
over a large redshift interval ($\beta >> 1$). 
In both cases, the age of the Universe at $z=6$ increases marginally.  

In fact, we can be even more general and consider that the Hubble parameter 
is some $H(z)$ for $z_1<z<z_2$, and equal to the Hubble parameter of 
$\Lambda{\rm CDM}$ at all other redshifts. 
Assuming $z_1<z_2<z_{\rm dec}$, the change in the age compared 
to the pure $\Lambda{\rm CDM}$ model is 
\be{yht2}
\Delta t(z_1)=\int_{z_1}^{z_2} {dz\over 1+z} \Big({1\over H(z)}-{1\over H_{\Lambda{\rm CDM}}(z)}.
\Big)
\ee
Since $1/(1+z)\leq 1/(1+z_1)$, we have 
\be{yht3}
\Delta t(z_1)< \frac{1}{z_1+1}\Omega_{\rm m}^{-1/2}H_0^{-1}\Delta{{\cal R}}.
\ee
For example, for $z_1=6$, with $h=0.7,\ \Omega_{\rm m}=0.3,\ \Delta{\cal R}=0.062$, 
this gives the upper limit
\be{yht4}
\Delta t(z=6)< 0.22\;{\rm Gyr}.
\ee

In terms of the linear growth factor, general constraints are not as 
straightforward to derive as now the exact features of the loitering 
phase play a crucial role. The linear growth rate in terms of the scale 
factor with
$H=X H_{\Lambda CDM}$ is
\be{lingrowa}
D''+\Big({3\over a}+{H_{\Lambda CDM}'\over H_{\Lambda CDM}}+{X'\over X}\Big)D'
-\frac 32 {H_0^2\over H_{\Lambda CDM}^2}{\om\over a^5}{1\over X^2}D=0,
\ee
where $'\equiv d/da$. Since loitering occurs at high redshifts, we can well
approximate $H^2_{\Lambda CDM}\approx H^2_{EdS}=H_0^2\om/a^3$. 
Approximating the loitering phase by $X(z)\approx X_0$ 
one finds that during this phase, $D\sim a^\beta$, with 
$\beta=(-1+\sqrt{24/X_0^2+1})/4$. For $X_0<1$, i.e. when the universe is 
loitering, $\beta$ is larger than one indicating faster growth than in 
the \lcdm (EdS) model. Quantifying the effect requires detailed knowledge about 
the loitering phase, making the age constraint, Eq. (\ref{yht4}),
a more robust constraint on any flat loitering model.

\section{Conclusions}
We have considered flat loitering universe models, both in the specific 
context of braneworld scenarios, and in general.  While increasing 
the time the universe spends at high redshifts and enhancing the 
linear growth factor might have desirable consequences for, e.g.,
modeling of the quasar population, modifications of the behavior of 
the Hubble parameter at high redshifts lead to changing the size
of the sound horizon at recombination. As this quantity is well
measured by the position of the peaks in the CMB angular power spectrum,
only a modest change is possible. Hence, a substantial increase in the age 
or linear growth factor at, say, $z=6$ are not allowed.
Only a modest increase in both quantities is possible, indicating that 
if a much older universe (or enhanced growth factor)is needed to accommodate 
observations, another new cosmological scenario is required.

\acknowledgments
{\O}E and DFM acknowledge support from the Research Council of Norway 
through project number 159637/V30.  In addition, {\O}E thanks NORDITA 
for hospitality.  TM is grateful to the Institute of Theoretical 
Astrophysics, Oslo, for hospitality.



\end{document}